# Improving the FAIRness and Sustainability of the NHGRI Resources Ecosystem

Larry Babb[1], Carol Bult[2†], Vincent J. Carey[3], Robert J. Carroll[4], Benjamin C. Hitz[5], Chris J. Mungall[6†], Heidi L. Rehm[1,3†], Michael C. Schatz[7†], Alex Wagner[8], NHGRI Resource Workshop Community*

[1] Broad Institute of MIT and Harvard, Cambridge, MA
[2] The Jackson Laboratory, Bar Harbor, ME
[3] Massachusetts General Hospital, Boston, MA
[4] Vanderbilt University Medical Center, Nashville, TN
[5] Stanford University, Stanford, CA
[6] Lawrence Berkeley National Laboratory, Berkeley, CA
[7] Johns Hopkins University, Baltimore, MD
[8] Nationwide Children's Hospital, Columbus, OH

* See Appendix B for the full list of Community Members
[†] Correspondences: carol.bult@jax.org, cjmungall@lbl.gov, hrehm@mgh.harvard.edu, mschatz@cs.jhu.edu

# Executive Summary

In 2024, individuals funded by NHGRI to support genomic community resources completed a Self-Assessment Tool (SAT) to evaluate their application of the FAIR (Findable, Accessible, Interoperable, and Reusable) principles and assess their sustainability. By collecting insights from the self-administered questionnaires and conducting personal interviews, a valuable perspective was gained on the FAIRness and sustainability of the NHGRI resources. The results highlighted several challenges and key areas the NHGRI resource community could improve by working together to form recommendations to address these challenges.

The next step was the formation of an Organizing Committee to identify which challenges could lead to best practices or guidelines for the community. The workshop's Organizing Committee comprised four members from the NHGRI resource community: Carol Bult, PhD, Chris Mungall, PhD, Heidi Rehm, PhD, and Michael Schatz, PhD. In December 2024, the Organizing Committee engaged with the NHGRI resource community to refine these challenges further, inviting feedback on potential focus areas for a future workshop. This collaborative approach led to two informative webinars in December 2024, highlighting specific challenges in data curation, data processing, metadata tools, and variant identifiers within the NHGRI resources.

Throughout the workshop planning process, the four Organizing Committee members worked together to create and develop themes, design breakout sessions, and create a detailed agenda. The workshop's agenda was intentionally structured to ensure participants could generate implementable recommendations for the NHGRI resource community.

The two-day workshop was held in Bethesda, MD, on March 3-4, 2025. The challenges received from NHGRI resources were classified into four key categories, forming the basis of the workshop. The four key categories are variant identifiers, data processing, data curation, and metadata tools. They are briefly described below, with greater details on their challenges and recommendations in subsequent sections.

- **Metadata Tools**:
  While metadata is vital for capturing context in genomic datasets, its usage and relevance can vary by domain, making it difficult to standardize usage. While various methods exist for annotating and extracting metadata, incomplete or inconsistent annotations often result in ineffective data sharing and interoperability, further reducing data usability and reproducibility.

- **Data Curation**:

Curation of annotations for genomics data is critical for FAIR-ness. Scalable curation solutions are challenging because of the multiple components for curation, including harmonizing data sets, data cleaning, and annotation. The workshop focused on identifying which aspects of data curation could be streamlined using computational methods while considering the barriers to increased automation.

- **Variant Identifiers:**
  Variant identifiers are standardized representations of genetic variants, crucial for sharing and interpreting genomic data in research and clinical work. They ensure consistent referencing and enable data aggregation. Standardizing variant identifiers is difficult due to varied formats, complex data, and distinct environments for generating and disseminating data.

- **Data Processing:**
  Data processing is a necessary first step in a FAIR environment. As there are many variant workflows, streamlining this process will ensure greater accuracy, reproducibility, interoperability, and FAIRness, driving advancements in clinical research. The workshop focused on addressing these aspects with a key focus on improvements and best practices around data processing for an NHGRI resource.

Several recommendations were made throughout the workshop's interactive sessions with the resources' participants. While many recommendations were specific to data processing, data curation, metadata tools, or variant identifiers, they can be grouped into core recommendations addressing common challenges within the NHGRI resource community. These core recommendations highlight the key themes that emerged across sessions and are listed in the nine recommendations below.

1. Increase transparency to enable effective sharing/reproducibility (documenting, benchmarking, publishing, mapping)
2. Develop entity schema and ontology mapping tools (between models, identifiers, etc.)
3. Annotate tools using resources to increase findability and reuse (Examples: EDAM Ontology of Bioscientific data analysis and data management)
4. Use standard nomenclature and identifiers
5. Make workflows usable by researchers with limited programming expertise
6. Implement APIs to improve data connectivity
7. Present data in an interpretable manner, along with machine readability
8. Develop artificial intelligence/machine learning (AI/ML) methods for scaling curation processes
9. Assess the impact of resources using an independent group that can assess return on investment and impact to health and scientific advancement.

An additional key collaborative outcome was the development of [Appendix A](#), which outlines ongoing and future efforts, including additional workshops, webinars, and meetings through the listed events provided by the NHGRI resource community. We hope that these activities will enable further advances in the implementation of FAIR standards and continue to foster collaboration and exchange across NHGRI resources and the global community.

# Metadata Tools

## Introduction

Metadata provides context for data, allowing it to be combined and linked with other data. Without adequate metadata, it can be hard or impossible to reanalyze, interpret datasets, or to combine multiple datasets together for re-analysis. Metadata encompasses a wide variety of information – administrative information about a study, biological and chemical properties of samples and their origin, details about library and sample processing, and the outputs and settings of physical and instrumentation and computational workflows. The division between data and metadata is not a hard and fast one, and may be context-dependent. For example, from a genomics perspective, information about a sequenced tissue sample (*tissue type, disease state of the donor, drug treatments, age*) are often considered metadata, but from a clinical perspective, these might be considered the primary data. But in either case, capturing this information is vital for analysis, integration, and reuse. This should be captured *prospectively*, at the time of data generation. Currently, a lot of data labeling and cleaning applies metadata *retrospectively*, which introduces both potential for error, as well as duplication of effort and inefficiencies.

The need for accurate metadata grows with the increase of sophistication of genomics technologies and with the range of phenotypic information captured clinically and in the lab. Elucidating the mechanistic underpinnings of health and disease requires composing together these different modalities, and their respective metadata. The rise of AI presents opportunities in how these disparate datasets can be integrated into multimodal foundation models. These models demonstrate powerful abilities to encode higher order patterns in data from large volumes of unstructured genomic or textual data, but metadata is required in order to pre-process data, label it, or link paired samples across multimodal datasets.

AI also presents opportunities in assisting with the organization of metadata, and helping with the more mundane aspects of data wrangling and metadata curation. Large language models, particularly when used within an agentic framework, can be used as assistants within metadata curation workflows. However, care should be taken as generative AI can introduce hallucinations and inaccuracies, so having an expert in the loop is vital.

## Challenge 1: Coordinating and harmonizing multiple data models

"*The good thing about standards is that there are so many to choose from.*" – Andrew S. Tannenbaum

There is a wealth of standards and data models in use in biomedicine, covering multiple kinds of omics modalities, clinical observations and phenotypes, knowledge about variants and gene functions, to name but a few. The cross-disciplinary nature of a lot of NHGRI research means that there are frequently multiple overlapping standards that need to be woven together. It's often not clear to data generators which metadata standards should be used, and in some cases the developers of metadata standards may not be aware of overlapping standards.

Recommendations

- Maintain a registry of semantic artefacts (schemas, data models, data dictionaries), in order to make it easier to find and contribute to standards. This could also be coordinated with providing recommendations for data management sharing plans.
- Encourage metadata standards to be more modular and composable. Rather than having multiple overlapping standards that each broadly range over genomic, clinical, and biomedical data, make it easier to connect smaller more specialized standards. The OBO Foundry provides a good model for how to do this for ontologies.
- Mappings and cross-walks should be published alongside standards, to facilitate data integration, and to make the connections between different metadata elements more explicit.
- For data that is either too new or specialized to map to an existing community standard, follow best practices such as always providing a data dictionary for any metadata elements or measurements collected. Frameworks such as LinkML schema sheets and CEDAR metadata templates are lightweight and easy ways to provide these in a way that maximizes reuse.

## Challenge 2: Achieving best practice for using ontologies

Ontologies are collections of inter-related terms that represent some aspect of a domain or activity. One of the main uses of ontologies is to standardize categorical values, allowing for both standardization and roll-up of data. For example, when capturing single-cell data, if free text labels are used, then it is harder to combine datasets. Using identifiers for cell ontology terms allows not only standardization, but also transfer of annotations to broader categories (e.g. rolling up 'alveolar macrophages' to 'macrophages'). Ontologies can also be used to annotate metadata elements in a schema or Common Data Elements (CDEs).

One challenge here is that ontologies are frequently broader in scope than is required for any given purpose. This makes it challenging to use the ontology for annotation and analysis, as there may be many terms that are out of scope. Here the provision of value sets can help with this problem, but creating, sharing, and maintaining these can be challenging.

Another challenge is the fact that all ontologies are under-resourced, and struggle to keep up with a backlog of requests. This creates pain points for data submitters, as they need to return to their metadata after a term request has been fulfilled, or simply use a more general term. Adding terms to ontologies is not always straightforward, as care must be taken to avoid introduction of duplicate concepts, or concepts that are similar but in disconnected branches in the ontology, which adds to its complexity. Additional automated methods would be welcome to help free up ontology developers from lower level tasks.

### Recommendations

- Metadata standards should enforce the use of ontology terms over free text strings, where appropriate.
- At the same time, not everything can or should be forced into an ontology term. Sometimes the use of a high-level ontology term plus additional free-form text is the best combination.
- Support efforts to create training and best practice material on the use and maintenance of ontologies.
- Encourage publishing standard value sets for commonly used packages of ontology terms, and encourage the use of tools such as CEDAR that allow for easy selection and application of value sets.
- Use transparent open mechanisms such as GitHub issues and pull requests for submitting to and making proposed changes to ontologies.
- Develop ways to evaluate the accuracy of using agentic AI for assisting with maintaining core ontologies, and with reducing the lag in the creation of new vetted ontology terms.

## Challenge 3: Ensuring provenance and linking of data objects

Biomedical and clinical data is becoming increasingly interlinked. An individual research subject or model organism may be a source for multiple samples and measurements, across different time points; and each of those samples may be subdivided or combined, for different omics assays. Furthermore, the raw data coming from those assays may be processed into multiple interlinked data objects. As such, the following recommendations are made in order to reuse and combine this data.

#### Recommendations

- Assign persistent resolvable identifiers for all entities and data objects, following best practices for identifiers, ensure identifier prefixes are used consistently and registered in [bioregistry.io](bioregistry.io), and reuse existing identifiers for pre-existing objects.
- Ensure relevant metadata is associated with the most appropriate identifier.
- Ensure that identifiers are linked together in a provenance graph, with clearly defined edge labels, allowing the tracing from final processed data objects back to their biological sources. Use schemas and ontologies such as PROV and OBI to capture the transformations along the provenance graph.

### Challenge 4: Tools for capturing and validating metadata

Researchers frequently cite providing metadata as a major pain point. This is often because the existing tools do not facilitate easy capture and mapping of metadata, and often require manual and error-prone re-entry of data. General purpose tools such as Excel are frequently used, which can introduce data errors, and lack mechanisms for ensuring conformance to a schema. Additionally, not all metadata is equally valuable, and the kinds of metadata that are required or recommended should be continually reviewed.

#### Recommendations

- Encourage the use of tools such as CEDAR and DataHarmonizer that facilitate capture of metadata that balances a simple spreadsheet-like interface and conformance to standards.
- Develop tools that facilitate capture of metadata at the point of data generation, avoiding manual reentry of data later on.
- Explore and evaluate ways to integrate agentic AI into existing and new tools.

# Data Curation

### Introduction

Curation ensures that data, including biological metadata annotations, comply with FAIR (Findable, Accessible, Interoperable, and Reusable) data management principles, thus enabling accessibility and reuse of data and knowledge by researchers, educators, policymakers, and the public. Curation promotes data integrity, sharing, reproducibility of research, compliance with legal and ethical standards, and knowledge discovery. Curation processes stand to benefit from disruptive AI technologies such as Large Language Models. Curation is also critical in making genomic data "AI ready".

Among the major challenges that hinder effective data curation are a lack of adherence to annotation and nomenclature standards in publications and in descriptions of data submitted to repositories, benchmarks for measuring and comparing the effectiveness of curation practices, leveraging new technologies to enhance the scalability of curation, and the sustainability of funding for curation.

## Challenge 1: Standards for publication that facilitate curation efficiency

Implementing annotation and nomenclature standards for publications and datasets facilitates curation efficiency and scalability. Such standards streamline curation processes and enhance downstream integration and interoperability of data across information systems and organisms. Unfortunately, even well-established nomenclature and annotation standards are not adhered to by authors and are not uniformly enforced by journals or data repositories. Mapping concepts and entities to annotation standards post publication is inefficient and time-consuming. Further, while significant resources are devoted to generating data, the funding of curation lags behind the need (Challenge 4).

Recommendations

- In partnership with grantees, publish and maintain an on-line catalog of NHGRI-funded projects that develop, maintain, and/or enforce genome data annotation and nomenclature standards.
- In partnership with grantees and experts in the field, regularly review and publish guidelines for annotation standards relevant to NHGRI-funded research programs.
- Require NHGRI-funded investigators to include the following in their data management plans:
    - details regarding the use of persistent identifiers for relevant data entities and
    plans for adherence to official nomenclature standards when they exist,
    - details about the ontologies and/or terminologies that will be used for annotations in data submissions and publications,
    - descriptions about how the data and annotations are related to NHGRI-funded databases and/or knowledge bases,
    - a data life cycle diagram/description.
- Fund and promote workshops for the research community on using appropriate standards for data representation in publication.
- Encourage practices that promote efficient downstream curation, data integration, aggregation, and retrieval.

## Challenge 2: Establishing useful benchmarks and standards for curation processes and database interoperability

Standards and benchmarks for data curation and database interoperability are needed to ensure that data can be easily shared, accessed, and reused by different user communities and
software platforms. In addition, benchmarks support meaningful evaluation and comparison of curation processes, helping identify areas for improvement and promoting data quality, interoperability, and alignment with FAIR principles. While curation benchmarks are dependent on the data type and research area, several recommendations can be implemented for most NHGRI-funded data resources to foster the development of common curation benchmarks and standards.

Recommendations

- Develop standard reporting metrics for curation processes, including measures related to data quality, curation scalability, and false positive and false negative results for curation methods.
- Require NHGRI-funded data resources to publish their curation processes and procedures in a standard format to promote meaningful comparisons across resources.
- Develop an interoperability network diagram for NHGRI-funded resources showing how curated data in one resource relates to data in another.
- Assemble and make available training sets for common curation tasks to support benchmarking.
- For curation standards:
    - require >2 pieces of evidence to support an annotation with clear source attribution,
    - promote curation of negative results/annotations (i.e., gene X does NOT have function Y),
    - encourage use of time-stamps for annotations and updates to annotations and the removal of annotations when the source is a retracted publication or data set,
    - require resources to make it clear which curated annotations are the result of AI or other computational methods.
- Organize a workshop specifically on curation standards and metrics, perhaps in collaboration with the International Society for Biocuration.

# Challenge 3: Developing and using AI and ML to enhance scalability of curation

As data volume and complexity continue to increase, developing AI and Machine Learning (ML) approaches will be increasingly crucial for designing and implementing scalable data curation processes. To ensure reliability and transparency, it is essential to establish clear metrics for evaluating the quality of AI-driven curation and to ensure AI methods give proper attribution for the evidence used for annotations. Identifying which components of data curation are most amenable to AI/ML will be necessary for deciding how to allocate resources for method development. As AI is a rapidly evolving field, most effort at this point in time should be identifying standard approaches to measuring the effectiveness of AI and ML rather than on the recommendation of specific methods or software platforms.

Recommendations

- Work with the research community to develop training sets for specific curation tasks, including "secret" validation datasets for benchmarking.
- Implement a centralized benchmarking resource, organized by curation tasks, to promote reusability of methods and to identify curation processes that need additional research and development before AI and ML approaches can be used effectively.
- Develop reporting guidelines for how and when AI methods are used for curation to promote transparency.
- Develop best practices for expert review of AI-generated annotations to ensure data quality.
- Develop training materials and workshops to train curation staff and educate users on the strengths and weaknesses of curation using AI and ML technologies.
- Stand up a centralized GitHub repository or similar open access resource, to share
documentation, training materials, training sets, and software.
- Develop guidance for licensing of data sets, methods, and tools for AI and ML to support data curation.
- Devise a system of formal recognition for the diverse contributions (e.g., training set development, expert assessment of AI curation calls, AI methods development) required to develop robust AI methods for curation.

## Challenge 4: Sustainability of funding for expert curation

A major challenge for data curation is that while the scale and complexity of data generated by the global biomedical research community continues to increase, the funding for expert curation remains flat or is diminishing. The sustainability of data curation is essential as the types of data and access methods continuously evolve, requiring ongoing updates to remain relevant and usable. Loss of access to critical data and tools can have catastrophic consequences for research continuity and reproducibility. Furthermore, the research community's work is deeply dependent on the accuracy, stability and availability of these curated resources, creating an interdependent ecosystem that relies on long-term support. AI/ML methods can only succeed when trained with accurate data resulting from careful expert manual curation.  Additionally, training and education in data science and related fields often depend on freely accessible resources, making sustained curation vital for building future expertise. While a solution for sustainability of curation is not a challenge unique to NHGRI there are areas in which NHGRI can make meaningful contributions.

Recommendations

- Work within NIH, across agencies within the U.S. and abroad, and through public-private partnerships, to promote diversification of funding for curation

- activities.
  - Identify data types that are a priority for NHGRI, make those priorities public, and regularly revisit them.
  - Develop a common strategy for data preservation for NHGRI-funded resources.
  - Provide funding for minimal maintenance activities for priority data sets and open-source software tools related to data curation.
  - Encourage the diversification of data storage and software packages among NHGRI-funded investigators.
  - Encourage and fund collaborative/communal development of curation tools and platforms to ensure broad dissemination and impact.
  - Encourage reviewers of grants with significant curation activities to emphasize impact over innovation.

## Conclusion

Curation is a fundamental activity for ensuring that genome data conform to FAIR data principles. Although data curation methods and practices have benefited from continuous, incremental improvements over the years, the emergence of Artificial Intelligence, Machine
Learning, and Large Language Models are disruptive technological forces that have the potential to transform how expert curation is performed.

The core recommendations outlined above are particularly important when funding resources are limited, to ensure that the investment in genomic data generation can be realized as advances in knowledge about the complex biological processes that underlie human health and disease. The implementation of the recommendations may be different for open-access versus controlled-access data types. To evaluate the return on investment that curation provides to the biomedical research community, NHGRI should consider contracting with an independent evaluation group to perform a detailed assessment every 3 to 5 years of the activities, outputs, outcomes, and impact of the curated data resources funded by NHGRI funds using methods employed to assess the economic and scientific value of resources such as those previously used by the European Bioinformatics Institute (EBI) (https://www.embl.org/documents/wp-content/uploads/2021/10/EMBL-EBI-impact-report-2021.pdf).

# Variant Identifiers

## Introduction

Variant identifiers are a key property of how knowledge regarding genomic variation is tracked, interpreted, and communicated across genomic resources. In the context of FAIR data practices, such identifiers should be machine-readable and provide human-interpretable descriptions, with clearly defined semantics and traceable provenance. We reviewed common systems for variant identification from genomic resources participating in the workshop, which focused primarily on small variants (insertions, deletions, and substitutions) and contextual aggregations of these.

Genomic resources represent these types of variants under three broad categories of variant identifiers:

1. **Registered Identifiers**
   Created by centralized resources or databases, registered identifiers such as ClinVar Variation IDs, dbSNP reference SNPs (rsIDs), and ClinGen Allele Registry IDs are opaque (randomly assigned) and must be resolved at the authoritative source to be understood. Such identifiers may reference entries that aggregate multiple related variant contexts (e.g., a set of individual alleles at a location comprising an rsID SNP).

2. **Computed Identifiers**
   Generated from the intrinsic properties of a variant (such as location, sequence change, and reference context), these identifiers may also be opaque, but are also deterministic and can be recalculated independently. The GA4GH Variation Representation Specification (VRS) is a leading framework in this category, supporting federated, cross-resource variant identification.

3. **Nomenclature-Based Identifiers**
   These include variant descriptions written in standardized syntax (e.g., HGVS, SPDI, stringified VCF seq-pos-ref-alt). Though infrequently used as formal identifiers, variant nomenclatures are ideal for human readability and are frequently used in literature and clinical reporting.

Due to their differing purposes and formats, identifier systems across these three categories are often used in parallel and even linked together. However, the complexity and lack of coordination between them poses serious challenges to data interoperability and reusability. We reviewed five key challenges with the use of variant identifiers and created key recommendations for genomic data resources to address these challenges to improve the FAIR sharing of variant data.

## Challenge 1: Understanding disparate variant concepts across systems

Variant identifiers are often used in isolation, and may lack sufficient context to interpret their meaning. For example, the use of an opaque registered identifier such as clinvar:7109 (a ClinVar Variation ID) provides limited information about the nature of the variant (where it occurs, what the change is). This is mitigated by accompaniment of the ID with a transparent representation (e.g., an HGVS or SPDI description), though variant nomenclatures alone may be insufficient for representing the complexity of variant aggregations like ClinVar Variation IDs. This is because such aggregations represent not only one or more genomic variants, but also any transcribed variant contexts as well as "liftover" representations of the variant on other genomic reference assemblies. Some resources have addressed this by using variant nomenclature concepts that precisely define these aggregations, such as the "Canonical SPDI" used by ClinVar. However, use of HGVS and SPDI together in a resource creates other potential points of confusion due to differences in the conventions underlying these nomenclatures, as discussed in [Challenge 2](#).

We also recognized that the use of opaque variant identifiers in variant reporting and inter-system exchange can lead to issues in concept findability, particularly if the variant identifier is a number or numeric string. For example, the top hits for a web search for the ClinVar variant concept "7109" returns addresses, product numbers, legislation, and a human gene. However, when namespaced appropriately (as in our above example of clinvar:7109), the correct concept becomes readily findable.

Recommendations

- Resources should clearly define the structure and meaning of concepts associated with variant identifiers, through use of external standards (e.g. GA4GH VRS, Canonical SPDI) or resource documentation.

- Variant concepts with opaque identifiers should use standardized nomenclature (e.g. HGVS descriptions) and resolvable URLs (e.g. https://reg.genome.network/allele/CA321211) to support provenance of the identified variant concept.

- When using opaque identifiers, provide in a namespace-prefixed format such as a CURIE (e.g., clinvar:7109, dbsnp:rs121909098) to improve findability.

- Register identifier prefixes with namespaced identifier resolver services like [identifiers.org](identifiers.org).

## Challenge 2: Understanding conventions underlying variant nomenclatures used by resources

When using variant nomenclatures, resources may deviate from existing community standards, leading to confusion in the interpretation of the variants as described by the resource. For example, some of the resources represented at this community meeting included nomenclatures using HGVS, SPDI, and hyphen-delimited VCF as seq-pos-ref-alt. Other resources use syntaxes derived from components of HGVS but are not wholly compatible with HGVS recommendations, such as "fragment" descriptions like c.299A>T that imply a reference sequence described elsewhere.

Even when using standardized nomenclatures, those discussed in this report follow different conventions for coordinate systems and variant normalization. When used together in a resource (e.g. HGVS alongside SPDI), some differences are noticeable on careful inspection but may lead to errors when users do not routinely account for such differences. These challenges are compounded when standard nomenclatures are indistinguishable from the syntax of resource-specific nomenclatures, which may potentially lead to misinterpretation of variant descriptions across resources. Moreover, some of these syntaxes (SPDI, stringified VCF) rely on additional assumed understanding of symbols for reference sequence, using shorthand such as "1" or "chr1" to represent sequences instead of sequence accessions such as refseq:NC_000001.11. These ambiguities risk conflation of, e.g., reference sequences from GRCh38 with prior reference assemblies.

Recommendations

- Resources should provide information on each component of variant description syntax. This MAY be through reference to an external nomenclature (e.g. HGVS v21.1).

- The coordinate system used (i.e. residue or inter-residue) should be indicated, as should the meaning of any numeric values in the nomenclature (e.g. position, sequence length). Resources should clearly indicate differences between coordinate system conventions, especially when they are likely to be misunderstood (e.g. SPDI's use of an inter-residue position).

- The reference sequences used for representing variants should be unambiguously described by the nomenclature, or accompanying metadata should be provided to disambiguate compact sequence representations such as "10" or "chr10".

- Variant normalization conventions, e.g. positional representation of ambiguous insertions or deletions in repeating sequence regions, should be described by the resource.

# Challenge 3: Translating across systems using different conventions for variant representation

When searching genomic resources for information, users may use different variant identifiers derived from a clinical research study, or may search by characteristics of variants such as genomic position or range. When users provide variant identifiers for searching a resource, the conventions used by their native system may differ from those displayed by the resource. This can create challenges for resource users, who may miss relevant data from the resource if it does not have sufficient support for the queried variant identifier system. For example, a user querying gnomAD v4 for "1-11114338-A-G" using the gnomAD stringified VCF conventions will be immediately understood, including the implied GRCh38 reference used by the v4 interface. That same query in other systems such as ClinVar is automatically identified as a gnomAD stringified VCF and through inference, is translated into a query of "NC_000001.11:11114337:A:G"[Canonical SPDI]. Other systems like the ClinGen Allele Registry require users to select their variant syntax before searching, and include support for gnomAD v4 identifiers. However, other mainstream resources, such as the UCSC Genome Browser, do not recognize the gnomAD resource syntax, and searches using this will fail.

This challenge is compounded by the complexity of supporting systems with different underlying normalization conventions, and a genomic range search is provided. Genomic resources following VCF-like conventions, for example, will "left-shift" insertions and deletions in repetitive sequence regions where the insertion site is ambiguous. In contrast, resources following HGVS conventions will "right-shift" such variants, and resources following VRS or SPDI conventions will represent these variants using "full-justification", avoiding arbitrary left- or right- shift considerations. Genomic range searches that are not aware of underlying resource conventions may not overlap relevant variant data that are left- or right- shifted outside of the query. Similarly, genomic coordinates in residue-coordinate systems have different meanings depending on whether the associated variant represents an insertion, deletion, or delins event with respect to a reference sequence, and variant event type is a nuance that is not often expressed in genomic range queries.

### Recommendations

- Resources should be able to accept and translate between common variant identifier systems (e.g. HGVS, VRS, VCF), and internal formats.

- Resources should specifically provide support for inter-residue coordinates and full-justification variant representation for genomic range queries to maximize search sensitivity.

## Challenge 4: Enabling new community variant representation standards in established genomic resources

Encouraging adoption of new or updated standards is challenging across communities with prior solutions to variant identification. For example, the National Marrow Donor Program uses HLA nomenclature to represent the diversity of alleles represented at the Human Leukocyte Antigen locus. Similarly, the pharmacogenomics community uses a "star allele" system for representing specific haplotype designations with associated "reference-agree" regions. When these systems are represented using HGVS, important contextual information can be lost or the expressions can become overly complex or unwieldy.

This challenge is compounded by the misuse of existing standards for new uses that they are not designed to handle. For example, clinical labs that have historically used HGVS for sequence variants will use the HGVS tandem duplication syntax to represent a copy number gain of a region, even if that copy gain is not necessarily in tandem (e.g., as determined by a CNV microarray). These tandem duplication expressions are submitted to other resources such as ClinVar, and are mis-represented as tandem duplications due to a lack of expressivity of the standard.

### Recommendations

- Coordinate with key members of the community (e.g. HGVS Variant Nomenclature Committee, GA4GH Genomic Knowledge Standards Work Stream) and key resource providers to understand needs and engage on solutions.

## Conclusion

While the recommendations described above provide a useful template for how resources can improve FAIR sharing practices with variant identifiers, two significant gaps remain that will require additional community investment. First, there is a clear need for tools that can reduce the barriers to supporting multiple community standards as a "plug-and-play" solution for legacy resources. Development of these common tools will help reduce duplication of effort to meet these recommendations, and drive collaboration among community participants to ensure these tools meet a diverse array of needs. Second, there is a need for resource providers that have challenges with "off-the-shelf" standards for variant representation to work in a community setting to improve inter-resource interoperability. Several such communities exist, most notably the GA4GH and HUGO, which together maintain most of the common variant identifier standards in use by human genomics resources today. Other community variant standards would benefit from coordinated development in these fora.

# Data Processing

## Introduction

Genomic research generates massive volumes of complex, heterogeneous data that must be processed through sophisticated computational workflows to yield meaningful biological insights. As sequencing technologies, multi-omics & imaging platforms, and other biomedical technologies continue to evolve, researchers require increasingly robust, scalable, and interoperable data processing systems. These systems must support cutting-edge analyses—from variant calling and transcript quantification to epigenomic annotation and machine learning—and also ensure that data and methods remain accessible and adaptable across different tools, institutions, and use cases.

Despite the major data production advances, the field continues to face substantial challenges with reproducibility. Even technologies that are supposed to assist with reproducibility and accessibility can fall short of their goals. For example, a recent analysis of over 27,000 Jupyter notebooks linked to articles in PubMed Central found that only 3% produced identical results when automatically rerun, largely due to missing documentation of software dependencies and inconsistent computational environments (Samuel S, Mietchen D. Computational reproducibility of Jupyter notebooks from biomedical publications. *Gigascience.* 2024. PMC10783158). This reproducibility crisis undermines trust in scientific findings and highlights the urgent need for improved standards, tooling, and infrastructure that can make genomic data processing more reliable, transparent, and repeatable. The current peer-review system rarely verifies computational reproducibility, suggesting an opportunity for automated validation tools and broader adoption of workflow best practices.

To address these issues, many challenges must be addressed, especially these three core challenges:

1. **Making analysis findable and reusable**
   This challenge requires standardized metadata, persistent identifiers, containerized tools, and registries to ensure that workflows can be discovered, cited, and reapplied reliably.

2. **Making analysis accessible and interoperable**
   This challenge involves building interfaces, APIs, and data standards that support a broad range of users—from novice biologists to expert bioinformaticians—and work seamlessly across platforms and institutions.

3. **Performing analysis over very large datasets**
   This challenge demands scalable, cloud-native, and/or federated solutions that can handle petabyte-scale data securely and cost-effectively, while enabling cross-cohort analyses and minimizing data movement.

Together, these challenges define the roadmap for building a robust, FAIR-compliant ecosystem for genomic data processing.

## Challenge 1: Making analysis findable and reusable

As genomic research scales in complexity and volume, the ability to find and reuse existing analysis workflows is essential for accelerating discovery, improving reproducibility, and reducing redundancy. Unfortunately, method descriptions and workflows are often buried in supplemental materials, scattered across repositories, or lack the metadata needed for meaningful reuse. This fragmentation makes it difficult for researchers to identify appropriate workflows for their datasets, to compare methods, or to build upon previous work. Moreover, even when workflows are accessible, they are frequently tied to specific environments or lack version control, limiting their portability and sustainability.

A findable and reusable analysis ecosystem would enable researchers to quickly locate vetted workflows, assess their provenance and applicability, and adapt them to new data with minimal friction. It would also promote the scientific rigor of computational methods, which are often only superficially described in natural language within publications. By assigning persistent identifiers, enforcing metadata standards, and embracing containerization, we can move toward a more reliable and discoverable ecosystem of reusable tools and workflows.

Recommendations

- **Integrate into the existing bioinformatics & workflow ecosystem**
  To ensure visibility and usability, analysis tools and workflows should be published in community-supported platforms such as GitHub, Dockstore, Bioconda, Galaxy, and Bioconductor. These ecosystems promote open source development, foster version control, and provide integration with package managers and workflow engines that make installation and execution easier across diverse systems.

- **Use persistent identifiers and versioning**
  Assigning persistent identifiers—such as DOIs via Zenodo or versioned

GitHub releases—ensures that workflows can be cited, tracked, and reused precisely. This practice supports reproducibility and scholarly attribution while enabling researchers to distinguish between stable releases and newer, potentially experimental versions of software.

- **Adopt software packaging and deployment systems**
  Distributing software via package managers like PyPI, Conda, and CRAN facilitates consistent installation of dependencies and helps standardize environments across users and systems. These platforms support automated testing, validation, and version pinning, all of which are critical for long-term usability and reproducibility.

- **Standardize workflow metadata and ontologies**
  Enhancing workflows with structured metadata—such as EDAM ontology tags—makes them more discoverable in registries and searchable through platforms like Dockstore or WorkflowHub. Natural language descriptions can be augmented or generated using AI to further improve indexing and accessibility, especially for non-specialist users.

- **Apply software engineering best practices**
  Continuous integration and deployment (CI/CD), test suites, code linting, and automated documentation checks are essential to keeping workflows functional and robust. Tools like GitHub Actions and Travis CI can automatically validate code with each update, reducing errors and promoting long-term maintenance. Open source licensing and clear documentation further support reuse.

- **Invest in training and documentation infrastructure**
  Effective reuse depends on users understanding how workflows function and how to adapt them. Projects should include layered documentation—from conceptual overviews and tutorials to full parameter references. Living documentation, such as Bioconductor vignettes or "cookbook-style" guides, enables workflows to evolve while remaining approachable.

- **Support generalizable and portable workflows**
  Workflows should be designed to move seamlessly across computing environments, whether local, HPC, or cloud. This includes the use of containers (e.g., Docker, Singularity) and support for multiple workflow languages. Broader adoption of standards like GA4GH Tool Registry Service (TRS) and Workflow Execution Service (WES) will also help workflows scale and interoperate effectively.

- **Leverage large language models for automation**
  LLMs (Large Language Models) can play a growing role in validating

documentation, generating tutorials, and identifying gaps in metadata. They can also assist in automatically producing screenshots, README files, or example uses, further lowering the barrier for workflow adoption and reuse.

Remaining Needs

While these recommendations offer a pathway to improving findability and reusability, major challenges remain. Support for newer workflow engines like Snakemake, NextFlow and full integration with metadata schemas such as RO-Crate are still under development. There is a persistent need for sustained maintenance funding, technical staff, and user training resources to ensure long-term viability. Finally, new tools and standards must be designed with interoperability and accessibility in mind, paving the way for broader adoption of these reusable resources across the genomics community.

## Challenge 2: Making analysis accessible & interoperable

Ensuring that genomic data analysis is accessible and interoperable is critical to enabling broad participation, seamless collaboration, and integration across diverse platforms and communities. Many researchers—especially those without formal training in bioinformatics or access to specialized infrastructure—face steep barriers to using cutting-edge tools. Meanwhile, inconsistencies in data formats, workflow interfaces, and authentication mechanisms limit the ability to integrate tools across institutional and national boundaries. Making analysis more accessible ensures that researchers of all backgrounds and institutions can contribute, while interoperability ensures that their tools and data can work together in a cohesive and scalable manner.

As the scientific community increasingly embraces distributed and cloud-based computing, accessible and interoperable systems must meet the needs of both human users and computational agents (such as AI systems). This includes building workflows that support graphical interfaces, integrating documentation and training, and developing standardized APIs that allow tools to interact programmatically. These efforts not only lower barriers for individual researchers but also enable more powerful, automated, and cross-disciplinary research through platform-agnostic integration.

Recommendations

- **Develop and publish APIs for core tools and workflows**
  APIs make it possible for researchers and automated systems to access data, trigger analyses, and retrieve results in a standardized way. Interactive tools, such as notebooks and GUIs, should expose API endpoints for data access and workflow control. Where possible, serverless architectures like Lambda functions can enable lightweight and scalable services without the need to manage infrastructure.

- **Publish data snapshots for reproducible and robust batch analyses**
  APIs alone are not sufficient when network reliability, version drift, or data mutability pose risks to reproducibility, especially for datasets and databases that can change over time. Static data snapshots—complete with provenance metadata and version control—are particularly important for clinical and regulatory applications, where reproducibility is paramount.

- **Harmonize datasets to be AI-ready**
  Interoperable systems require data to follow well-defined schemas, with consistent metadata and clear provenance. Validated, "tidy" datasets enable automated tools—including LLMs and ML pipelines—to parse and analyze information effectively. Programmatic validators should ensure consistency across datasets and flag anomalies automatically.

- **Adopt and contribute to community standards**
  Standards such as the GA4GH APIs, OAuth for authentication, and FastAPI or Swagger for documenting interfaces ensure that tools can be used across platforms and integrated into broader ecosystems. Harmonizing around shared specifications fosters compatibility while reducing the need for custom code and workarounds.

- **Design for users with limited programming expertise**
  Platforms like Galaxy and tools with no-code/low-code interfaces should be integrated into workflows to support users with varying skill levels. Natural language interfaces, such as AI-driven chatbots, can further lower barriers to entry, enabling researchers to launch analyses, explore results, and troubleshoot issues with conversational prompts.

- **Ensure compliance with regional and institutional policies**
  Policies like the U.S. FISMA/FedRAMP (NIST 800-171, NIST 800-53) and the EU Cyber Resilience Act introduce security and privacy requirements that must be met by tools and platforms. Interoperable systems must incorporate secure data access controls, auditable provenance, and architecture that can adapt to regulatory constraints.

- **Create registries and documentation for APIs and services**
  Just as Dockstore catalogs workflows, a registry of APIs—especially for core NHGRI resources—would enhance discoverability and reuse of services. A service registry should document every API endpoint, inputs, and outputs using community standards like EDAM, and facilitate credential management and access across platforms. This will also promote the use of LLMs through MCP servers and other emerging technologies.

### Remaining Needs

While significant progress has been made, several gaps persist in making analysis truly accessible and interoperable. Universal standards for workflow portability are still evolving, and cloud-specific dependencies complicate cross-platform execution. Many interfaces remain overly complex, particularly for non-specialist users, and the integration of tools across credentialed, protected systems remains cumbersome. Ongoing efforts must prioritize streamlining authentication, developing user-friendly interfaces, harmonizing API documentation, and securing sustainable funding for engineering and support roles. Achieving widespread accessibility and interoperability will require not only technical solutions but also community engagement, policy coordination, and targeted investment in infrastructure.

## Challenge 3: Analysis over very large datasets

Modern genomic research increasingly depends on analyzing large-scale datasets, ranging from terabytes to petabytes in size. These include population-scale cohorts, multi-modal datasets (genomics, transcriptomics, imaging, clinical records), and cross-consortia integrations such as All of Us, the UK Biobank, and AnVIL. Processing such vast and diverse data volumes requires scalable, parallelized, and cost-effective infrastructure—often beyond the capabilities of local computing resources. As a result, cloud-based execution and federated analysis models have emerged as practical and necessary solutions for enabling high-throughput discovery while minimizing data movement, improving security, and managing costs.

Large-scale analysis also presents practical and technical challenges across the full lifecycle of research: from data ingestion and storage, to compute orchestration, to results dissemination and interpretation. Efficient data structures, workflow automation, and harmonized APIs are essential to make such systems usable and reliable. Moreover, equitable access to computing infrastructure remains a key concern, as many research groups—particularly those in resource-constrained environments—lack sufficient cloud credits, technical training, or system support to leverage these capabilities fully.

### Recommendations

- **Use clusters or clouds to run workflows in parallel**
  Scaling analysis across thousands of samples requires access to distributed computing environments that support task parallelization. This includes HPC clusters, cloud-based platforms, and hybrid architectures that can dynamically scale capacity ("cloudbursting") as needed. Support for GPUs and TPUs is increasingly important for AI-powered workflows and high-throughput pipelines.

- **Prefer federated analysis over bulk data transfers**
  Instead of moving large datasets across institutions or platforms—which is costly, slow, and often restricted—federated analysis enables computation to occur where the data resides. This approach reduces egress fees, limits duplication, and strengthens privacy protections. It also allows multiple institutions to contribute to joint analyses without relinquishing data control.

- **Adopt efficient data formats and compression strategies**
  To maximize performance and minimize storage needs, large genomic datasets should be stored and processed using modern formats such as Apache Parquet for tabular data and sparse VCFs for variant calls. Streaming data structures and sketching algorithms can support lightweight, approximate analyses without the overhead of full dataset downloads.

- **Establish a robust ecosystem of tools for managing big data**
  Successful large-scale analysis depends on a surrounding ecosystem of utility tools for ingestion, transformation, filtering, visualization, and export. This includes support for OLTP/OLAP databases, scalable file systems, and parallel query engines that can handle both transactional and analytical workloads efficiently.

- **Invest in workflow optimization and automation**
  Running workflows at scale requires automated mechanisms for task scheduling, caching, and resource optimization. Tools such as Nextflow Tower, Snakemake Profiles, Galaxy Cached results, and Cromwell Call-Caching reduce redundancy and improve efficiency. Automated cost modeling tools can help researchers select between on-prem and cloud options based on workload characteristics.

- **Provide equitable access to computing infrastructure**
  Programs like NIH STRIDES, NSF ACCESS, and CZI open science credits should be expanded to provide cloud resources to under-resourced researchers and institutions. Training programs must also support users in setting up, executing, and scaling their workflows in cloud and federated environments.

- **Streamline access to protected datasets across platforms**
  Many critical datasets are gated behind complex access processes. Emerging tools like DUOS (Data Use Oversight System) and "meta access applications" for GRU-governed datasets help consolidate permissions and approvals. APIs should allow programmatic access while respecting data use restrictions, enabling scalable yet secure federation.

Remaining Needs

Despite strong advances in cloud computing and federated infrastructure, several barriers still hinder routine large-scale analysis. Equitable access remains a core issue, especially for early-career researchers and those outside major consortia. Workflow parallelization and optimization are not yet fully automated or intuitive. Federated systems need better standardization and robust security protocols, and access to protected datasets remains fragmented across platforms. Continued investment in tooling, training, infrastructure, and policy frameworks is needed to ensure that scalable data analysis is achievable not just in theory, but in practice, for all members of the genomics community.

## Conclusion

Enabling robust, scalable, and reproducible data processing in genomics hinges on addressing three interconnected challenges: making analysis findable and reusable, accessible and interoperable, and scalable for very large datasets. Each requires a mix of technical infrastructure, community standards, and sustained investment in documentation, training, and engineering. As genomic datasets grow in size and complexity, and as research becomes increasingly collaborative and cross-platform, there is a pressing need for workflows that are well-documented, modular, cloud-ready, and governed by FAIR principles.

Looking ahead, AI and large language models will likely play an increasingly prominent role in accelerating workflow documentation, automating quality control, and aiding tool discovery, while their integration raises new challenges related to reliability, transparency, and validation, particularly in clinical and regulatory contexts. Ensuring these AI systems are "research-grade" and trustworthy will require ongoing evaluation, clear benchmarking, and tight coupling with standardized metadata and ontologies. If the genomics community can collectively rise to these challenges, the result will be a more equitable, efficient, and innovation-friendly research ecosystem capable of transforming data into insight at a global scale.

# Overall Conclusion

The aforementioned recommendations are intended to serve as guidelines for the NHGRI resources. The overall goal of the workshop was to bring together members of the NHGRI resource Community to collaborate and address shared challenges within genomics research.

Participants engaged in thoughtful dialogue, identified key challenges, and contributed to developing these recommendations. NHGRI is collaboratively considering different aspects of these recommendations for implementation.

# Next Steps

There are several ways that the NHGRI community can continue to work together to achieve FAIRness. Future efforts include additional workshops, continuing webinars, meetings through the collaborative forums listed in [Appendix A](Appendix A), and continued conversations with deeper dives to solve these and additional challenges. A potential future example could be discussing FAIRness maturity as described through the Joint Research Centre data policies. NHGRI resource community members are encouraged to take these recommendations forward, building upon them and driving the next phase of FAIRness.

# Appendix

## Appendix A: Collaborations within NHGRI Resources

There are numerous opportunities to collaborate within the NHGRI resource community. These include community forums, open office hours, and additional collaborative events. Below is a comprehensive list in alphabetical order of these opportunities compiled by NHGRI resource community members who attended the March 2025 workshop.

**Alliance Community Forum:**
- ➢ A place for discussion related to the Alliance of Genome Resources
- ➢ https://community.alliancegenome.org/categories

**Alliance Genome Calendar of Events:**
- ➢ The Alliance of Genome Resources hosts webinars and other events for the genomic research community. Unless otherwise specified, all Office Hours are held at Noon Eastern Time. Most events will be approximately 30 minutes, including time for questions. Links to recordings of past events are listed on the page.
- ➢ https://www.alliancegenome.org/event-calendar

**AnVIL Calendar of Events:**
- ➢ The NHGRI Genomic Data Science Analysis, Visualization, and Informatics Lab-space (AnVIL) hosts a variety of events to support users and foster collaboration. These include monthly AnVIL Demos, which feature live demonstrations of platform capabilities and scientific analyses, followed by Q&A and user support. The AnVIL Champions program empowers individuals to serve as local resources for their communities and connect them with the AnVIL team. AnVIL also convenes its user community annually at the AnVIL Community Conference. Users can engage year-round through the AnVIL Community Forum at https://help.anvilproject.org and can explore additional events at anvilproject.org/events.
- ➢ https://anvilproject.org/events

**Bioconductor conference archive**
- ➢ The annual Bioconductor conference has a conventional URL of the form biocYYYY.bioconductor.org. Most talks are recorded and made available as YouTube videos on Bioconductor's YouTube channel.
- ➢ https://bioc2023.bioconductor.org/schedule/

➢ has video links for most talks from 2023, for example, this talk concerns Bioconductor's interface to The Carpentries. A playlist for the 2024 conference is here.

**Bioconductor support site**
➢ Collect 15+ years of questions and answers about computational genomic data science with Bioconductor.
➢ https://support.bioconductor.org

**Bioconductor workshop archive**
➢ To support authoring, teaching, and learning about topics in computational genomic data science, this Galaxy-based system provides access to pre-loaded Rstudio/posit workbench instances that guide students through workshops on single-cell transcriptomics, spatial transcriptomics, ontology, package development, and other themes.
➢ https://workshop.bioconductor.org

**ClinGen - Clinical Genome Resource**
➢ The Clinical Genome Resource, or ClinGen, is a National Institutes of Health-funded initiative to increase the community's knowledge about the relationship between genes and health. How to get involved
➢ https://www.clinicalgenome.org/

**Cytoscape**
➢ Cytoscape is an open-source software platform for visualizing complex networks and integrating them with any type of attribute data. The links below provide ways to learn about, contribute to, and interact with the Cytoscape community.
➢ Contributors: https://cytoscape.org/documentation_developers.html
➢ Training: https://tutorials.cytoscape.org/
➢ Automation: https://automation.cytoscape.org/

**Federated Variant-Level Matching (VLM) Project**
➢ The VLM project aims to support a global federated network of genomic databases, powered by GA4GH standards.
➢ https://www.ga4gh.org/what-we-do/ga4gh-implementation-forum/federated-variant-level-matching-vlm-project/

**GA4GH Cloud Work Stream calls**
- ➢ Biweekly meetings coordinating standards for executing, sharing workflows, and sharing data on the cloud
- ➢ [https://www.ga4gh.org/work_stream/cloud/](https://www.ga4gh.org/work_stream/cloud/)
- ➢ [https://github.com/ga4gh/wiki/wiki](https://github.com/ga4gh/wiki/wiki)

**GA4GH Genomic Knowledge Standards (GKS) Workstream**
- ➢ Designs technical standards that enable computer-to-computer exchange of information about genomes, improving the search and application of genomic knowledge.
  Standardizes the exchange of genomic knowledge through common APIs, schemas, and software. Enables interoperability between diagnostic laboratories, electronic health records, researchers, and knowledge bases — leading to better health outcomes.
- ➢ [https://www.ga4gh.org/work_stream/genomic-knowledge-standards/](https://www.ga4gh.org/work_stream/genomic-knowledge-standards/)

**Galaxy Project Events**
- ➢ Purpose- The Galaxy Project offers a wide range of events to foster community engagement and collaboration, including the annual Galaxy Community Conference, the Galaxy Webinar Series, Developer Round Table meetups, CollaborationFests, Papercuts CoFests, and regular Community Calls. Additionally, the Galaxy Training Academy provides globally accessible training resources and events, reaching thousands of analysts worldwide.
- ➢ [galaxyproject.org/events](galaxyproject.org/events)
- ➢ [https://training.galaxyproject.org/training-material/events/index.html](https://training.galaxyproject.org/training-material/events/index.html).

**Galaxy Project Series**
- ➢ Some Galaxy events are part of a series.
- ➢ [Galaxy Webinar Series](Galaxy Webinar Series)
- ➢ [Galaxy Developer Round Table meetups](Galaxy Developer Round Table meetups)
- ➢ [CollaborationFests and Papercuts CoFests](CollaborationFests and Papercuts CoFests)
- ➢ [Galaxy Community Calls](Galaxy Community Calls)

**GENCODE webinars and helpdesk**
- ➢ The human and mouse GENCODE gene sets are the primary data sets that the Ensembl outreach and training team uses to demonstrate features of the Ensembl Genome Browser. This acts as an opportunity for users to learn more about GENCODE and all the secondary analyses we provide through the Ensembl Genome Browser. This information is delivered through live webinars, archived training material and workshops at large conferences.
- https://training.ensembl.org/
- Queries on GENCODE are handled via our helpdesk
- gencode-help@ebi.ac.uk

**gnomAD forum**
- This is a public forum for users to post questions (for the gnomAD team and other users), report bugs, and provide feedback on gnomAD.
- https://discuss.gnomad.broadinstitute.org/

**GWAS Catalog**
- The GWAS Catalog is a freely available database of human genome-wide association studies. We deliver in-person training sessions and webinars via EMBL-EBI training, and have a number of routes of contact, listed below and at https://www.ebi.ac.uk/gwas/docs/about
- A **public forum** for discussion with the GWAS Catalog team and other users: gwas-users@ebi.ac.uk  Send a "Please subscribe me email" to join this list.
- **Training resources:** https://www.ebi.ac.uk/training/search-results?query=gwas&domain=ebiweb_training&page=1&facets=
- **Helpdesk:** gwas-info@ebi.ac.uk
- Report bugs and start technical discussions via GitHub: https://github.com/EBISPOT/goci/issues
- **Announcement List:** We announce new releases and other developments on our announcement list, gwas-announce@ebi.ac.uk
-  You can also follow us on LinkedIn, BlueSky @gwascatalog.bsky.social ,and X @GWASCatalog

**Knowledge Portal Focus Group**
- This is an intimate session where we will meet with key members of the community to gather feedback and show new features from the A2FKP. This is held **every 2 months on the second Thursday from 12-1 PM ET** and alternates with our Knowledge Portal Webinars.
- https://a2f.hugeamp.org/news.html
- Join the call using the following Zoom
- https://broadinstitute.zoom.us/j/99741357044?pwd=JzdySEADAz8P0UddielTAYG6Ya6ltL.1
- Passcode: 130805

**Knowledge Portal Webinars**
- ➢ The Knowledge Portal Webinars are designed to educate our users and introduce them to new portal features, community research driven by our knowledge portals, and guest speakers to gather direct feedback from the community. This is held **every 2 months on the second Thursday from 12-1 PM ET** and alternates with our Knowledge Portal Focus Groups.
- ➢ https://a2f.hugeamp.org/news.html
- ➢ https://broadinstitute.zoom.us/j/95039888063?pwd=wz3HFcER6Lwc2NhTRPV8t77nAuvQrM.1
- ➢ Passcode: 363662

**LinkML Community Calls**
- ➢ Monthly community meetings to discuss tools and modeling practices for capturing metadata for resources.
- ➢ https://linkml.io/linkml/faq/contributing.html

**OBO Academy Seminar Series**
- ➢ Engage in education and discussion on the use of ontologies for annotation data (genomics and other types).
- ➢ https://oboacademy.github.io/obook/courses/monarch-obo-training/

**PGS Catalog**
- ➢ The PGS Catalog is a freely available database of human genome-wide association studies. We deliver in-person training sessions and webinars via EMBL-EBI training
- ➢ https://www.ebi.ac.uk/training/search-results?query=pgs&domain=ebiweb_training&page=1&facets=
- ➢ The PGS Catalog can be contacted via the helpdesk  pgs-info@ebi.ac.uk
- ➢ Github https://github.com/PGScatalog/

**PhenX Toolkit**
- ➢ The PhenX (consensus measures for **Phen**otypes and e**X**posures) Toolkit (https://www.phenxtoolkit.org) is a freely available web-based catalog of measurement protocols recommended for research with human participants. The Toolkit is an established resource that includes protocols across 30 broad research domains (e.g., Obesity and Rare Genetic Conditions) and six focused collections (e.g., Blood Sciences Research, Substance Use, Use Disorders, and Recovery). Use of PhenX protocols facilitates data sharing and cross-study analyses and can increase the impact of individual studies. Below are links to the PhenX Toolkit, workshops, tutorials, and how to join the PhenX listserv (includes opportunity to provide feedback on Working Group preliminary protocols being considered for inclusion in the Toolkit). Additionally, the PhenX team hosts a booth at the American Society for Human Genetics annual meeting to interact and engage with the research community.
- ➢ PhenX Toolkit: https://www.phenxtoolkit.org/
- ➢ PhenX Workshops: https://test.phenxtoolkit.org/news/workshops
- ➢ PhenX Newsletters and link to join listserv (newsletters, release announcements, Working Group community outreach): https://test.phenxtoolkit.org/news/newsletters
- ➢ PhenX Tutorials: https://www.youtube.com/channel/UCbDroMNmfIMwCA1fa3masyw

**Reactome**
- ➢ Outreach takes on a variety of forms, such as public talks, lectures, visiting universities and colleges, and supporting traditional science events (meetings, conferences, and workshops). Education is a major focus of Reactome community outreach and aims to inform biologists, clinicians, and bioinformaticians about what Reactome can do. We undertake many community outreach events each year to educate our user base. We also provide several self-paced educational resources. We advertise them on our website and social media. We also make 1-on-1 meetings available to users via a calendar booking link.
- ➢ https://reactome.org/community/events
- ➢ https://www.ebi.ac.uk/training/search-results?query=reactome&domain=ebiweb_training&page=1&facets=https://reactome.org/community/outreach

**UCSC Genome Browser**
- ➢ The UCSC Genome Browser is a free interactive map viewer for annotations on the human chromosomes and thousands of other genomes. The user can choose between hundreds of colored annotation "tracks", switch them on and off, zoom and move around, highlight, filter and export the view as tables and PDFs.
- ➢ We collaborate in one way or another with almost every genomics data resource. We import annotations from more than 100 public projects, mostly from NIH (many of them listed in this document, e.g., UniProt, ClinGen, ClinVar, GnomAD, Reactome, GWAS Catalog, and many others), and a few commercial databases. Packages in Bioconductor can load our annotations and other genome browsers, e.g. IGV, rely on our database. Users can upload their own data and display it on our website and all other browsers using our "Track Hub" technology, and we then copy some of these to our servers when they are popular or important for NIH.

- ➢ Some users contribute code to our website via Github, e.g., to display their own data in a better way, or they describe to us how they want their data to be shown, and we implement their suggestions.
- ➢ **Outreach:** We describe updates bi-monthly on our announcement mailing list, on Bluesky, Twitter and Linkedin, we run in-person workshops for > 1500 participants/year, reply daily to a dozen questions at [genome@soe.ucsc.edu](genome@soe.ucsc.edu), provide interactive tutorials in our "Help" menu and a Youtube channel at [https://www.youtube.com/c/ucscgenomebrowser](https://www.youtube.com/c/ucscgenomebrowser)
- ➢ Link: [https://genome.ucsc.edu](https://genome.ucsc.edu) (and mirrors [genome-asia.ucsc.edu](genome-asia.ucsc.edu) and [genome-euro.ucsc.edu](genome-euro.ucsc.edu), for reliability and faster access)

**UniProt**
- ➢ UniProt is the world's leading high-quality, comprehensive, and freely accessible resource of protein sequence and functional information. We import data and cross-reference out to many of the listed resources and undertake many outreach and training events each year to increase and educate our user base. These are all advertised on our website and social media platforms, and links to many of these can be found at the link given below. An annual strategic partners meeting helps guide the direction of future activities and support collaborative activities:
- ➢ [https://www.ebi.ac.uk/training/search-results?query=uniprot&domain=ebiweb_training](https://www.ebi.ac.uk/training/search-results?query=uniprot&domain=ebiweb_training)
- ➢ [www.uniprot.org](www.uniprot.org)

**WashU Epigenome Browser Workshop**
- ➢ (Before the pandemic) We used to host interactive workshops that would introduce genomic resources produced by consortia, including ENCODE, Roadmap Epigenomics, and 4DN, and the WashU Epigenome Browser and associated tools (https://epigenomegateway.wustl.edu/). The Epigenome Browser hosts hundreds of thousands of genomic datasets produced by large consortia and investigators, and supports navigation of the data and its interactive visualization, integration, comparison, and analysis. Attendees will gain hands-on experience with exploring the most current epigenomic resources and with advanced visual-bioinformatics tools, including gene set view, genome juxtaposition, and chromatin-interaction display, inventions unique to the WashU Epigenome Browser. The workshops will demonstrate the power of "big bio-data" through specific examples.
- ➢ [https://epigenomegateway.wustl.edu/support/](https://epigenomegateway.wustl.edu/support/)

**Working group on standards for reporting predicted effector genes**

- ➢ A collaborative workshop hosted by the Knowledge Portal Network and GWAS Catalog, to define the standards, infrastructure, and incentives required to promote and enable sharing & interoperability of predicted effector gene lists.
- ➢ [https://kp4cd.org/2024_PEG_workshop](https://kp4cd.org/2024_PEG_workshop)

# Appendix B: NHGRI Resource Workshop Attendees List

| Name - First | Name - Last | Resource(s) representing |
|---|---|---|
| Joanna | Amberger | OMIM |
| Carolyn | Applegate | OMIM |
| Lawrence | Babb | ClinGen, GA4GH |
| Elspeth | Bruford | HGNC & VGNC |
| Carol | Bult | Mouse Genome Database, Alliance of Genome Resources |
| Vincent | Carey | Bioconductor |
| Robert | Carroll | AnVIL Clinical Resource |
| Ishwar | Chandramouliswaran | NIH |
| Katherine | Chao | Genome Aggregation Database (gnomAD) |
| Chuming | Chen | UniProt |
| Christopher | Churas | Cytoscape |
| Jyoti | Dayal | Program Director for PhenX |
| Peter | D'Eustachio | Reactome, Path2GO |
| Valentina | Di Francesco | National Human Genome Research Institute, NIH |
| Idan | Gabdank | National Human Genome Research Institute, NIH |
| Marc | Gillespie | Reactome |
| Anupama | Gururaj | ImmPort |
| Maximilian | Haeussler | UCSC Genome Browser |
| Tong | Hao | Protein protein interaction |
| Laura | Harris | GWAS Catalog, PGS Catalog |
| David | Hill | GENCODE |
| Ben | Hitz | IGVF RegulomeDB ENCODE |
| Wayne | Huggins | PhenX Toolkit |
| Edward | Huttlin | BioPlex Project |
| Julie | Jurgens | Knowledge Portal Network, AMP-CMD, Common Fund Data Ecosystem |
| Nicolas | Keller | National Human Genome Research Institute, NIH |
| Natalie | Kucher | NHGRI AnVIL |
| Melissa | Landrum | ClinVar |
| Jonathan | Lawson | Broad Institute |
| Daofeng | Li | WashU Epigenome Browser, IGVF, HPRC, Washington University, Saint Louis |
| Renan | Martin | GeneMatcher and VariantMatcher, Johns Hopkins University |
| Fergal | Martin | GENCODE |
| Stephen | Mosher | AnVIL |
| Chris | Mungall | GO, Alliance |
| Erin | Ramos | NIH |
| Jordan | Ramsdell | BioTeam |
| Heidi | Rehm | ClinGen, gnomAD, GA4GH |
| Michael | Schatz | AnVIL, Galaxy |
| Lynn | Schriml | Human Disease Ontology Knowledgebase (DO-KB) |
| Stephan | Schurer | NHGRI MorPhiC resource |
| Alan | Scott | OMIM |
| Alessandra | Serrano Marroquin | National Human Genome Research Institute, NIH |
| Reed | Shabman | NIAID Data Ecosystem Discovery Portal |
| Neethu | Shah | Clinical Genome Resource |
| Michael | Sinclair | Sylvester Data Portal (SDP) |
| Cynthia | Smith | Mouse Genome Informatics |
| Lincoln | Stein | WormBase, AGR, Reactome |
| Jessica | StLouis | BioTeam |
| Courtney | Thaxton | The Clinical Genome Resource (ClinGen) |
| Sharna | Tingle | National Human Genome Research Institute, NIH |
| Dusica | Vidovic | NHGRI MorPhiC |
| Alex | Wagner | GA4GH |
| Chris | Wellington | National Human Genome Research Institute, NIH |
| Ryan | Whaley | PharmGKB |
| Michelle | Whirl-Carrillo | PharmGKB, CPIC, PharmCAT |
| Mark | Woon | PharmGKB |
| Jennifer | Wortman | BioTeam |
| Matt | Wright | Clinical Genome Resource (ClinGen) |
| Cathy | Wu | UniProt |
| Sandhya | Xirasagar | National Human Genome Research Institute, NIH |
| Andy | Yates | GA4GH |
| Denis | Yuen | Dockstore |

# Appendix C: Improving the FAIRness and Sustainability of the NHGRI Resources Ecosystem Agenda.

**Monday, March 3, 2025**
9:00 AM         Registration and Networking
9:30 AM         Welcome and Logistics
10:10 AM        Breakout Session A (Two Parallel Sessions):
                    Variant Identifiers, Data Curation
12:10 PM        LUNCH
1:10 PM         Networking
1:40 PM         Breakout Session B (Two Parallel Sessions):
                    Metadata Tools, Data Processing
3:50 PM         Open Discussion on Recommendations
5:00 PM         Adjournment Day 1

**Tuesday, March 4, 2025**
8:30 AM         Networking
8:50 AM         Welcome Day 2
9:00 AM         Breakout Session A (Two Parallel Sessions):
                    Variant Identifiers, Data Curation
                Breakout Session B (Two Parallel Sessions):
                    Metadata Tools, Data Processing
12:00 PM        LUNCH
1:00 PM         Plenary Readout of Workshop Recommendations
2:15 PM         Adjournment Day 2